\documentstyle[eqsecnum,aps]{revtex}
\newcommand{\bq}{\begin{equation}}
\newcommand{\enq}{\end{equation}}
\newcommand{\bqa}{\begin{eqnarray}}
\newcommand{\enqa}{\end{eqnarray}}

\begin{document}
\title{Photoluminescence Study of Carbon Nanotubes }
\author{H. X. Han,$^{(1)}$ \cite{byline}, G. H. Li,$^{(1)}$  W. K. Ge,$^{(2)}$  Z. P. Wang,$^{(1)}$  Z. Y. Xu,$^{(1)}$ S. S. Xie,$^{(3)}$  B. H. Chang,$^{(3)}$  L. F. Sun,$^{(3)}$, B. S. Wang,$^{(1)}$  G. Xu,$^{(4)}$  Z. B. Su$^{(4)}$}
\address{$^{(1)}$National Laboratory for Superlattices and Microstructures, Institute of Semiconductors, Chinese Academy of Sciences, Beijing 100083, China}
\address{$^{(2)}$Department of Physics, Hong Kong University of Science and Technology,Clear Water Bay, Kowloon, Hong Kong}
\address{$^{(3)}$Center for Condensed Matter Physics, Institute of Physics, Chinese Academy of Sciences, Beijing 100080, China}
\address{$^{(4)}$Institute of Theoretical Physics, Chinese Academy of Sciences, Beijing100080, China}
\date{\today}
\maketitle
\begin{abstract}
Multiwalled carbon nanotubes, prepared by both electric arc discharge and chemical vapor deposition methods, show a strong visible light emission in photoluminescence experiments. All the samples employed in the experiments exhibit nearly same super-linear intensity dependence of the emission bands on the excitation intensity, and negligible temperature dependence of the central position and the line shapes of the emission bands. Based upon theoretical analysis of the electronic band structures and optical transition, it is suggested that besides the electronic transitions across the fundamental gap, the transitions between $\pi$ and $\sigma$ conduction bands are the major source of the light emissions. A two-step transition mechanism is proposed.   
\end{abstract}

Carbon nanotubes (CNTs), since their discovery by Iijima\cite{Iijima}, have attracted ever-increasing interest because of their novel structures and potentials in quantum engineering. A single wall carbon nanotube (SWNT) can be envisaged as a seamless cylinder constructed by wrapping up a two-dimensional (2D) graphite sheet with a hexagonal network. Theoretical studies on the electronic properties predict that CNTs could be metallic or semiconducting depending sensitively on the tube diameter and the chirality of the hexagonal lattice\cite{Mintmire,Hadama,Saito,White}. The diameter and the chirality are uniquely determined by ${\bf C}_h = n{\bf a}_1+m{\bf a}_2{\equiv}(n, m)$, which connects two crystallographically equivalent sites on the 2D graphite sheet, where n, m are integers and ${\bf a}_1$, ${\bf a}_2$ the unit vectors of a 2D hexagonal lattice. The condition for the nanotubes to be metallic is $2n + m = 3q$, where q is an integer\cite{Saito}. In recent years, considerable progresses have been achieved in experimental investigations on the optical properties of CNTs\cite{Heer,Rao,Kasuya,Sugano,Pimenta,Bonard,Chen,Liu}. Many papers have reported Raman scattering studies on the vibrational properties of the tubular materials, including resonant transitions between one-dimensional (1D) electronic sub-bands, while less attention, to our knowledge, has been drawn to light emission properties up to date. In this letter, we report our photoluminescence (PL) studies of multiwalled carbon nanotubes (MWNTs) as well as the theoretical interpretation on the salient features of the PL bands. We observed a strong visible light emission in the PL experiments. The light emission bands are centered in a range from 2.05 eV to 2.3 eV for different MWNT samples and which are far beyond the fundamental gaps of semiconducting CNTs. In addition, the PL intensities show nearly the same super-linear dependence upon the excitation intensity, which is not consistent with a single photon excitation process. Based on the theoretical analysis\cite{Wang} of the electronic band structures, especially on the symmetry properties of both ${\pi}$ and ${\sigma}$ electronic states, a two-step transition mechanism is suggested. This mechanism interprets our experimental results very well.\par
We measured PL spectra of four MWNT samples. Samples A, B and C were prepared by electrical arc discharge (AD) method, while sample D by chemical vapor deposition (CVD) method. Samples A and D were purified by high temperature oxidation technique. In TEM measurements, small amount of nanoparticles are seen from the unpurified samples B and C, and sample D shows much longer tube length than those of the other samples. The details of sample growth have been described elsewhere\cite{Pan}. In the PL measurements, the tubular samples were excited by the 406.7 nm line or 325 nm line of a Coherent C100 Krypton ion laser and a HeCd laser, respectively. Light spots on the sample surface were about 10 $\mu m$. To reveal the excitation intensity dependence of the PL properties, laser power on the sample surface was varied from 5mW to 50 mW. Sample temperature was varied from 10 K to 300 K. Under the excitation, we observed a strong light emission from the tubular samples. The emission light was analyzed by a Jobin-Yvon HRD1 double grating monochromator and detected by a GaAs cathode photo-multiplier tube, then fed to computer through a photon counter. \par Fig.1 shows (a) room temperature PL spectra of CNT samples A, B and D excited by 406.7 laser line, and (b) PL spectra of sample C excited by  325 nm laser line at different temperature. It is shown in Fig.1 (a) that all the samples show a strong light emission band. The central position of emission bands at room temperature are extended from 2.05 eV to 2.2 eV, far beyond the fundamental band gap of semiconducting CNTs, for both purified and unpurified samples prepared by AD as well as for the one by CVD method. As shown in Fig.1 (b), the PL intensities decrease with increasing temperature. On the other hand, the emission bands show a Gaussian line shape with full width at half maximum (FWHM) of about 0.8 eV and no significant temperature effects on the band shapes as well as on FWHM. It is also noted that there is no remarkable temperature dependence on the central positions of the PL bands . The PL intensity dependence on the excitation intensity is depicted in Fig.2. All the samples exhibit nearly the same super-linear intensity dependence of the PL bands on the excitation intensity and which cannot be explained by a single photon excitation process. In is worth noting that the PL bands were highly reproducible in both peak shapes and positions as confirmed in our experiments on all the samples and from different sample locations. We also tested PL spectra of a highly oriented pyrolytic graphite (HOPG) sample,  no visible light emission was observed. All these features strongly indicate that the PL bands are associated with the electronic properties inherent to the CNT structures and not due to any extrinsic effects such as various carbonaceous materials in our samples.\par 
     We noticed an interesting report by Bonard et al.\cite{Bonard} who observed field emission induced luminescence, centered at about 1.8 eV, due to optical transitions between the defects levels at the top of the tubular structures. What we would like to address in this letter is the intrinsic optical properties of CNTs. A recent study on the electronic structure by Charlier et al.\cite{Charlier} presents theoretical results for the electronic density of state (DOS) of chiral CNTs. They find that the electronic band structures of CNTs with chiral symmetry are similar to those of zigzag and armchair ones. The metallic tubes show constant DOS near the Fermi energy due to linear dispersion while the semiconducting ones show moderate band gaps with sharp spikes of DOS at the band edges. For all the cases, however, the DOS show a series of spikes of Van Hove singularities. They also noted that in spite of different electronic band structures near the Fermi energy, the general feature of DOS apart about 1 eV from the Fermi energy is quite similar and that is in good agreement with recent experimental results\cite{Wildoer,Teri}. As has been shown in the resonant Raman scattering experiments\cite{Rao,Kasuya,Sugano,Pimenta}, the spikes in 1D DOS of the $\pi$ electrons can provide initial or final states, or both, for strong optical transitions. Charlier et al.\cite{Charlier} also pointed out that for the nanotubes with the diameter of 1.3 nm, there exist strong optical transitions (at about 1.7-2.0 eV) between the spikes of the full and empty ${\pi}$ bands. These arguments are in good agreement with the results of resonant Raman scattering experiments\cite{Rao,Kasuya,Sugano,Pimenta} and recent report\cite{Liu} of the absorption peak at 1.9 eV.  In our case, however, the central positions of the emission bands take the value well beyond the transition range mentioned above. Furthermore, the super-linear intensity dependence of the PL bands cannot be explained by simply assigning it to the transitions between occupied and unoccupied $\pi$ sub-bands since it would result in a linear intensity dependence of PL bands. All these arguments imply that the PL bands observed in our experiments cannot be characterized by ordinary discussed fundamental band gaps of the tubular structures and that we need to consider electronic states in a wider range.\par 
To understand the features of the observed light emission band, we have re-examined theoretically the electronic band structures and optical transition properties of CNTs. Since the electronic structure of multi-walled CNT can  fairly be approximated  by  superposition of the constituting monolayer CNT's\cite{saito2}, we only consider monolayer CNT here. In order to consider both ${\pi}$ and ${\sigma}$ valence and conduction bands in a simple and reliable way, we adopt $SP^3S^*$ model which is known to be capable of giving rise a substantial improvement for usual $SP^3$ model, especially for the lowest conduction bands\cite{Vogl}. By further applying the chiral symmetry-adopted coordinates\cite{White}, we calculated the band structures for several nanotubes with different chirality and depict some  results in Fig.3. In the present case, ${\sigma}$ and ${\pi}$ bands are still uncoupled (here the coupling between ${\pi}$ and ${\sigma}$ bands due to curvature is not considered\cite{Benedict} ). Compared with the usual $SP^3$ model, the results of the $SP^3S^*$ model calculations still keep the structure of the ${\pi}$ electron states unchanged, while the top of the ${\sigma}^+$ valence band and the bottom of the ${\sigma}^-$ conduction band are lowered. Particularly the lowest ${\sigma}^-$ conduction band is flattened, i.e. it produces a high peaks of DOS as shown in Fig. 3. It can be seen that the difference in chirality does not effect much the DOS feature at high energy range. This peak is located around 2.1 eV above the bottom of the lowest spikes of the ${\pi}^-$ electrons and will play an important role for the observed PL bands. The direct dipole excitation for ${\pi}^+$ valence electrons to states of the ${\sigma}^-$ conduction band is nearly forbidden, but a two-step transition process ${\pi}^+{\to}{\pi}^-{\to}{\sigma}^-$ is allowed. Because of the different symmetry properties,  electrons excited to the ${\sigma}^-$ band could hardly be relaxed to the ${\pi}^-$ band via phonon scattering process or via short-range part of Coulomb interaction. As a result, the electrons will be accumulated at the edge of the ${\sigma}^-$ band and then the electronic transition from the ${\sigma}^-$ band to the bottom of the ${\pi}^-$ bands generate the strong visible emission bands. Such a kind $\sigma$-$\pi$ electron transition plays an import role, as pointed by R. Saito et. al. \cite{saito2}. This two-step excitation process gives the dominant contribution to the super-linear intensity dependence of the PL bands. Such a picture implies the capability of the visible light emission from the carbon nanotubes.\par
     The involvement of the high-excited states, which play an important role in the above discussion, is also consistent with the recent report by Chen et al.\cite{Chen} on optical limiting of the CNTs. They observed strong optical limiting behavior with photon energy of 2.3 eV, which is in reasonable agreement with our theoretical analysis as well as with the super-linear intensity dependence of the PL band on the excitation intensity. If there is no such a high-excited state, optical absorption will show sub-linear dependence on the excitation intensity and finally lead to saturation behavior of the nanotubes. It is also reasonable that the light emission was not observed from HOPG because of the absence of the 1D spikes of the DOS in this layered structure. In addition, the polarization of the light resulted from the $\sigma$ - $\pi$ transitions are perpendicular to the surface, and then the emitted light could be re-absorbed in graphite. This argument can also provide a reason for no report on the observation of the PL bands in most carbon nanotube samples, which are bunched, while our samples are well separated from tube to tube and hence make light emission much easier.
\par 
As a conclusion, we present experimental evidence of visible light emission associated with the electronic properties of CNT structures. The emission bands show super-linear intensity dependence on the excitation intensity. Based on the theoretical consideration on the electronic states, a two-step transition process is suggested, and which explains our experimental results very well. The observation of strong light emission from CNTs as well as their strong non-linear properties suggests the potential of CNTs in quantum photonic device applications. Further investigations on the optical emission properties as well as more deliberate non-linear optical features are expected due to the existence of the high-excited states in the optical properties of CNTs mentioned in this letter. 
This work was partly supported by Natural Science Foundation of China.

\begin{figure}
\caption{(a) PL spectra of CNT samples A, B and D at room temperature and (b) PL spectra of sample C at different temperature.}
\end{figure}

\begin{figure}
\caption{Intensity dependence of PL band of sample A, B and D on the excitation intensity at 10 K.}
\end{figure}

\begin{figure}
\caption{ Calculated DOS of $\pi$ and $\sigma$ electrons of carbon nanotubes with different chirality. Insert show expanded DOS of $\pi$ and $\sigma$ electrons of the carbon nanotubes.}
\end{figure}

\end{document}